\documentclass[useAMS,usenatbib,usegraphicx,letterpaper]{mn2e}
\usepackage{times,amssymb,hyperref,subfigure,epsfig,aas_macros}

\begin{document}

\title[The Fomalhaut C debris disc]{Discovery of the Fomalhaut C debris disc}

\author[G. M. Kennedy et al.]{G. M. Kennedy\thanks{Email:
    \href{mailto:gkennedy@ast.cam.ac.uk}{gkennedy@ast.cam.ac.uk}}$^1$, M. C. Wyatt$^1$,
  P. Kalas$^2$, G. Duch\^ene$^{2,3}$, \newauthor B. Sibthorpe$^4$, J.-F. Lestrade$^5$,
  B. C. Matthews$^{6,7}$, and J. Greaves$^8$\\
  $^1$ Institute of Astronomy, University of Cambridge, Madingley Road, Cambridge CB3
  0HA, UK \\
  $^2$ Department of Astronomy, University of California, B-20 Hearst Field Annex,
  Berkeley, CA 94720-3411, USA \\
  $^3$ UJF-Grenoble 1 / CNRS-INSU, Institut de Plan\'etologie et d'Astrophysique (IPAG)
  UMR 5274, 38041 Grenoble, France \\
  $^4$ SRON Netherlands Institute for Space Research, NL-9747 AD Groningen, The
  Netherlands \\
  $^5$ Observatoire de Paris, CNRS, 61 Av. de l'Observatoire, 75014 Paris, France \\
  $^6$ National Research Council of Canada, 5071 West Saanich Road, Victoria, BC, Canada V9E 2E7\\
  $^7$ University of Victoria, Finnerty Road, Victoria, BC, V8W 3P6, Canada\\
  $^8$ School of Physics and Astronomy, University of St Andrews, North Haugh, St Andrews, Fife KY16 9SS, UK}
\maketitle

\begin{abstract}
  Fomalhaut is one of the most interesting and well studied nearby stars, hosting at
  least one planet, a spectacular debris ring, and two distant low-mass stellar
  companions (TW~PsA and LP~876-10, a.k.a. Fomalhaut B \& C). We observed both companions
  with \emph{Herschel}, and while no disc was detected around the secondary, TW~PsA, we
  have discovered the second debris disc in the Fomalhaut system, around LP~876-10. This
  detection is only the second case of two debris discs seen in a multiple system, both
  of which are relatively wide ($\gtrsim$3000 AU for HD 223352/40 and 158 kAU [0.77 pc]
  for Fomalhaut/LP~876-10). The disc is cool (24K) and relatively bright, with a
  fractional luminosity $L_{\rm disc}/L_\star = 1.2 \times 10^{-4}$, and represents the
  rare observation of a debris disc around an M dwarf. Further work should attempt to
  find if the presence of two discs in the Fomalhaut system is coincidental, perhaps
  simply due to the relatively young system age of 440 Myr, or if the stellar components
  have dynamically interacted and the system is even more complex than it currently
  appears.
\end{abstract}


\section{Introduction}\label{s:intro}

Fomalhaut is perhaps the most interesting nearby stellar and planetary system outside our
own. The primary hosts inner and outer dust belts with temperatures similar to the Solar
System's Asteroid and Kuiper belts \citep{2013ApJ...763..118S}, and in addition hosts an
enigmatic exoplanet, Fomalhaut b \citep{2008Sci...322.1345K}. The system also contains
two additional stars, a secondary TW~PsA \citep[Fomalhaut
B,][]{1997ApJ...475..313B,2012ApJ...754L..20M} and a tertiary LP~876-10 \citep[Fomalhaut
C,][]{2013AJ....146..154M}. At only a few dozen light years from Earth, this remarkable
system provides a unique laboratory in which to observe one outcome of star and planet
formation in detail \citep[see][for reviews of the Fomalhaut planetary and stellar
systems]{2013ApJ...775...56K,2013AJ....146..154M}

The \emph{Herschel} \citep{2010A&A...518L...1P}\footnote{\emph{Herschel} is an ESA space
  observatory with science instruments provided by European-led Principal Investigator
  consortia and with important participation from NASA.} DEBRIS Key Programme observed an
unbiased sample of nearby stars with the goal of discovering and characterising the
extra-Solar Kuiper belt analogues known as ``debris discs'' around hundreds of nearby
stars \citep[e.g.][]{2010A&A...518L.135M}. The sample comprises the nearest $\sim$90 each
of A, F, G, K, and M spectral types that are not confused by proximity to the Galactic
plane \citep{2010MNRAS.403.1089P}, and includes TW~PsA and LP~876-10 (both at 7.6
pc). Fomalhaut itself was observed as part of a guaranteed time programme
\citep{2012A&A...540A.125A}.

M dwarfs are rarely observed to have excesses at any IR or millimetre wavelength
\citep[e.g.][]{2007ApJ...667..527G,2006A&A...460..733L,2009A&A...506.1455L,2009ApJ...698.1068P},
so the detection of an IR excess around LP~876-10 is interesting as the discovery of a
rarely seen object. However, given that the Fomalhaut system is already known to host a
debris disc and at least one planet, this detection of yet another planetary system
component is particularly exciting. For example, a possible origin of the eccentric ring
and planet around Fomalhaut A is a previous encounter with LP~876-10, and the LP~876-10
disc may show signs of such an encounter. Below, we describe the \emph{Herschel}
observations and describe some basic properties of the debris disc around LP~876-10 and
quantify the non-detection of debris around TW~PsA, noting some possible implications of
this discovery for the overall system status and evolution.

\section{Observations}

\begin{figure*}
   \begin{center}
     \hspace{-0.25cm} \includegraphics[width=0.35\textwidth]{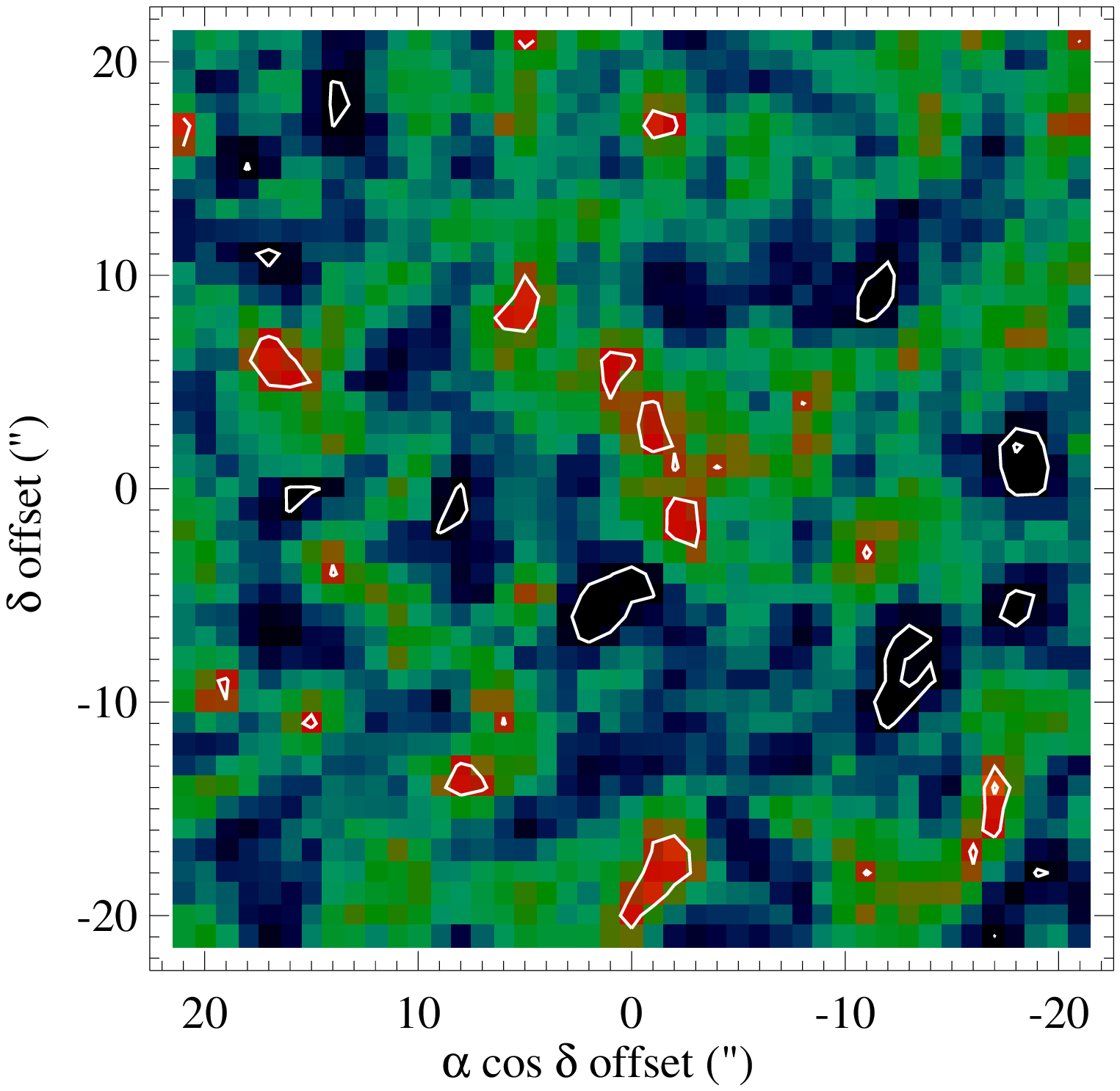} 
     \hspace{-0.5cm} \includegraphics[width=0.35\textwidth]{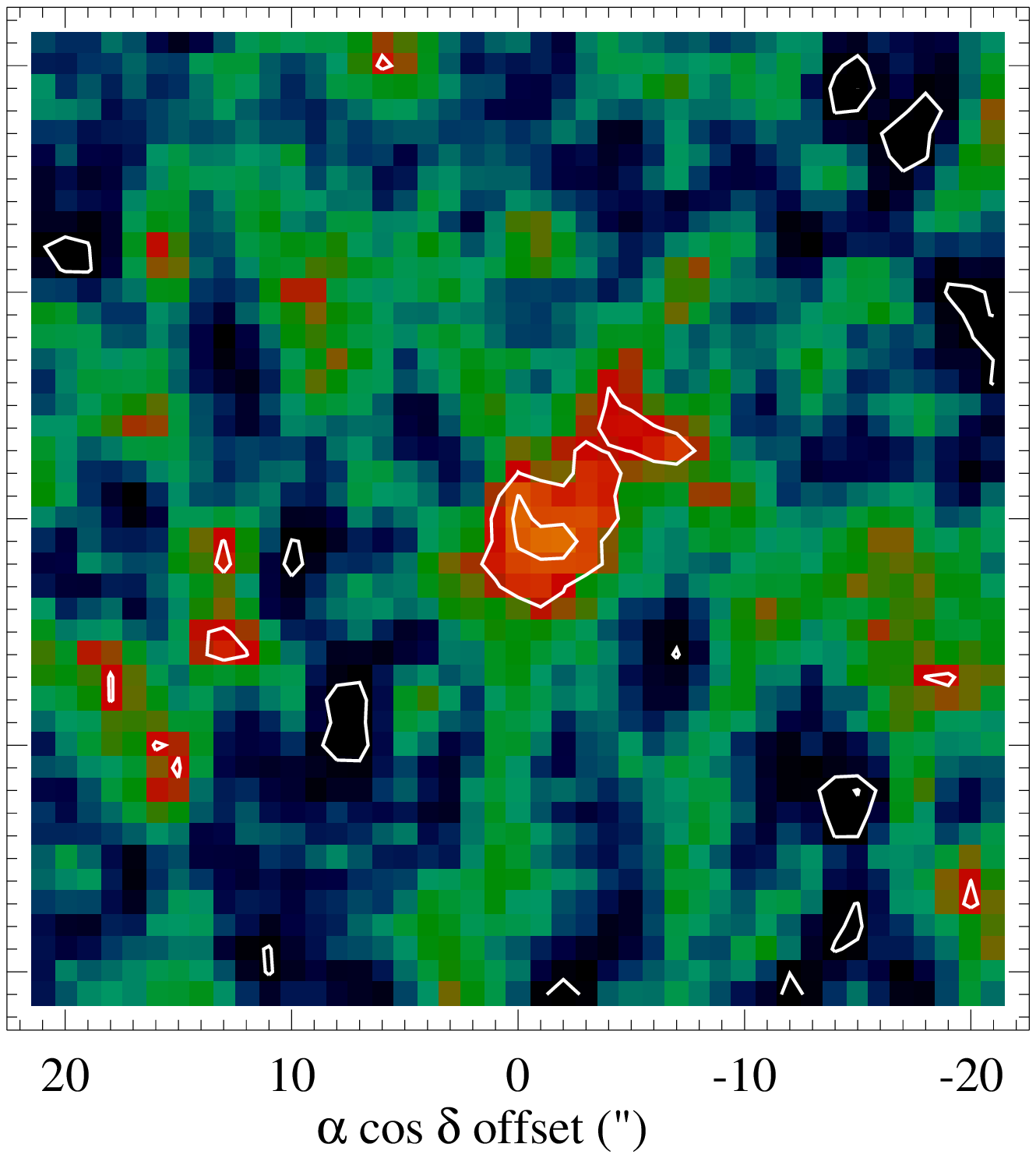} 
     \hspace{-0.5cm} \includegraphics[width=0.35\textwidth]{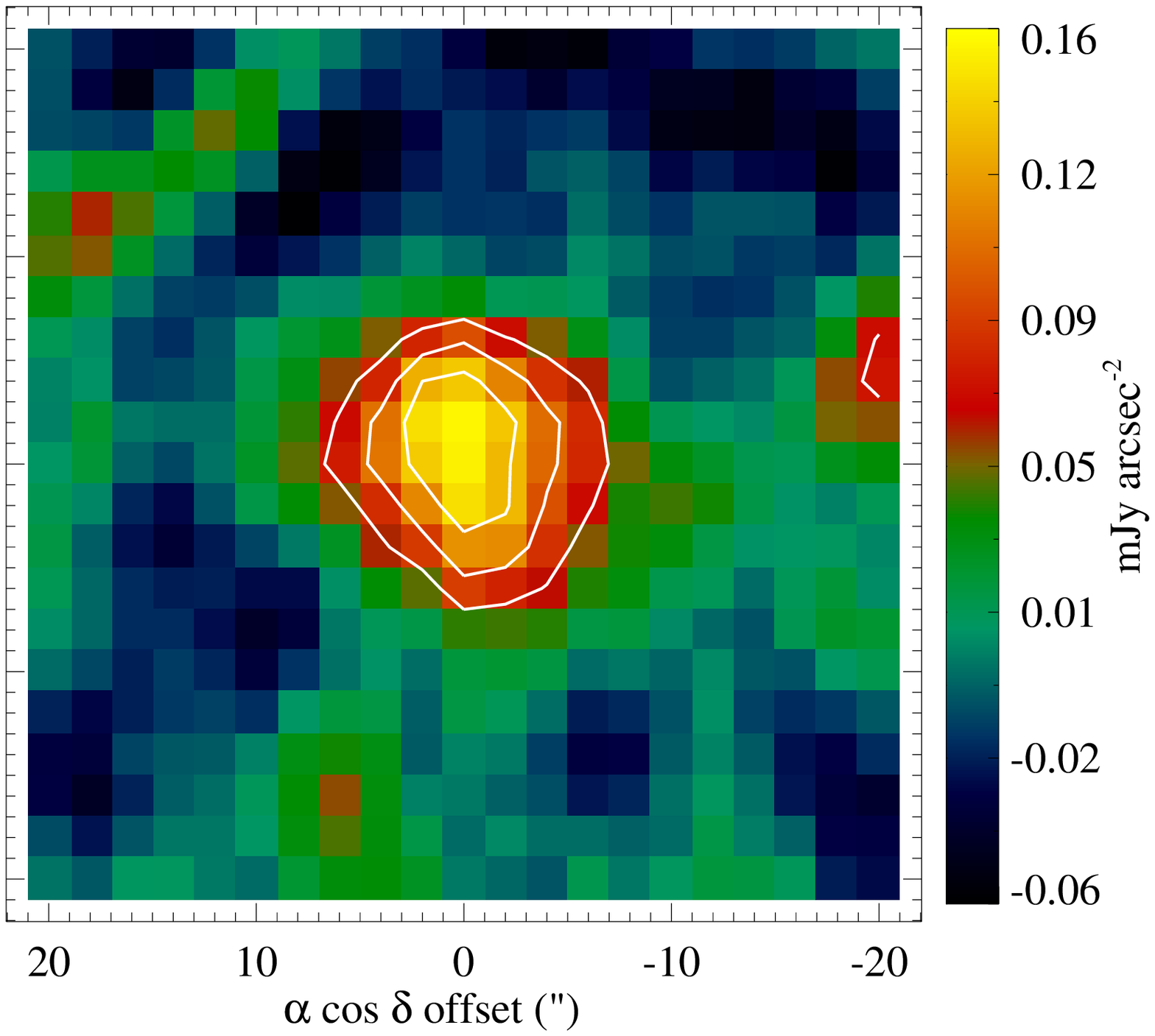} 
     \caption{\emph{Herschel} PACS images of LP~876-10, at 70, 100 and 160 $\mu$m from
       left to right. The expected December 2010 source position of $\alpha=22$h 48m
       4.7s, $\delta=-24^\circ$ 22m 9.9s is at the centre of the images, and the PACS
       pointing accuracy for these observations is about 2 arcseconds at
       1$\sigma$. Contours are drawn at intervals of $\pm 2, 3$ and
       4$\sigma$.}\label{fig:ims}
  \end{center}
\end{figure*}

\begin{table}
  \caption{\emph{Herschel} observations of TW~PsA and LP~876-10. PACS observes at 70 or
    100 $\mu$m, and always at 160 $\mu$m. Each ObsID represents a single scan direction, and the two differ by 40$^\circ$. The duration is for a single ObsID.}\label{tab:obs}
  \begin{tabular}{lllll}
    \hline
    Target & ObsID & Date & Instrument & Time (s) \\
    \hline
    TW PsA & 1342211140/1 & 13/12/2010 & PACS100 & 445 \\
    LP~876-10 & 1342211142/3 & 13/12/2010 & PACS100 & 445 \\
    LP~876-10 & 1342220645 & 8/5/2011 & SPIRE & 721 \\
    LP~876-10 & 1342231937/8 & 6/11/2011& PACS70 & 1345 \\
    \hline
  \end{tabular}  
\end{table}

TW~PsA and LP~876-10 were observed using the \emph{Herschel} Photodetector and Array Camera
\& Spectrometer \citep[PACS,][]{2010A&A...518L...2P} and Spectral and Photometric Imaging
Receiver \citep[SPIRE,][]{2010A&A...518L...3G} instruments (see Table \ref{tab:obs}). The
PACS observations used the standard ``mini scan-map'', which comprises two sets of
parallel scan legs, each taken with a 40$^\circ$ difference in scan direction. The SPIRE
observations were taken using a single standard ``small map'', with two scans at
near-90$^\circ$ angles. The raw timelines were projected onto a grid of pixels
(i.e. turned into images) using a near-standard HIPE pipeline
\citep{2010ASPC..434..139O}.

\subsection{LP~876-10 (Fomalhaut C)}

LP~876-10 was first observed at 100 and 160 $\mu$m in December 2010, and showed the
presence of a probable IR excess so additional PACS 70 and 160 $\mu$m and SPIRE 250, 350,
and 500 $\mu$m observations were also obtained. Compact emission from the expected
position of LP~876-10 was detected in the 100 and 160 $\mu$m PACS images, but not at 70
$\mu$m or in the SPIRE images. The PACS images are shown in Figure \ref{fig:ims}, with
emission clearly detected at 160 $\mu$m, marginally detected at 100 $\mu$m, and not
detected at 70 $\mu$m. The detection positions are consistent with the expected star
position given the 2 arcsec at 1$\sigma$ pointing uncertainty of \emph{Herschel}. Fitting
the emission with observations of the calibration target $\gamma$ Dra as a model point
spread function (PSF) shows that the emission is consistent with that from a point
source. Some structure is seen to the NW at 100 $\mu$m, though after PSF subtraction this
emission is only 1-2$\sigma$ significant.

We measured the PACS source fluxes using both PSF fitting and aperture photometry, for
which the results were consistent. The final PACS flux measurements are $3.3 \pm 1.5$
mJy at 70 $\mu$m (i.e. formally a non-detection), $7.8 \pm 1.9$ mJy at 100 $\mu$m, and
$15.5 \pm 2.8$ mJy at 160 $\mu$m. The SPIRE non detections are largely limited by
confusion noise, for which the 3$\sigma$ limits at 250, 350, and 500 $\mu$m are 15.9,
18.9, and 20.4 mJy (see SPIRE Observer's Manual). Emission of about 10 mJy/beam is present
at the expected stellar location in the 250 $\mu$m image, though the significance is
sufficiently low that the emission does not appear point-like, and we cannot be sure that
it is associated with LP~876-10 and not due to the background. We therefore set the
3$\sigma$ limit in this band to 25.9 mJy.

\begin{figure}
  \begin{center}
    \hspace{-0.25cm} \includegraphics[width=0.48\textwidth]{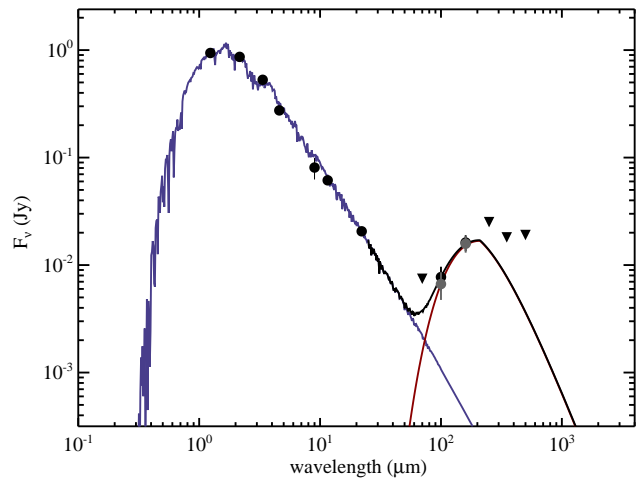}
    \caption{SED for LP~876-10. Dots are fluxes and triangles 3$\sigma$ upper
      limits. Black symbols are measured fluxes and grey symbols are star-subtracted
      (i.e. disc) fluxes. The 3200K stellar photosphere model is shown in blue, the 24K
      blackbody disc model in red, and the star+disc spectrum in black.}\label{fig:sed}
  \end{center}
\end{figure}

Figure \ref{fig:sed} shows the spectral energy distribution (SED) for LP~876-10. The
stellar photosphere is derived by fitting AMES-Cond atmosphere models to 2MASS, AKARI,
and WISE \citep{2006AJ....131.1163S,2010A&A...514A...1I,2010AJ....140.1868W} photometry
from 1.2 - 22 $\mu$m via least squares minimisation. The best fitting model has effective
temperature $T_{\rm eff} = 3200 \pm 140$ K and luminosity $0.0049 \pm 0.0004 L_\odot$,
consistent with values derived by \citet{2013AJ....146..154M}. The photospheric fluxes
are 2.2, 1.1, and 0.4 mJy at 70, 100, and 160 $\mu$m. Figure \ref{fig:sed} includes the
PACS photometry and SPIRE upper limits, showing clear ($>3\sigma$) far-IR excess emission
at 100 and 160 $\mu$m. To estimate the temperature of this emission we fitted a simple
blackbody model, finding a temperature of $24 \pm 5$K. We have ``modified'' the blackbody
disc spectrum, multiplying by $( \lambda_0/\lambda)$ beyond $\lambda_0=210 \mu$m
\citep{2008ARA&A..46..339W}, to account for inefficient long-wavelength emission by small
grains and ensure a more realistic prediction of the sub-mm disc brightness. This
modification is not required by the SPIRE limits however, and the true spectrum beyond
160 $\mu$m is uncertain. This model has a fractional luminosity $L_{\rm disc}/L_\star
\equiv f =1.2 \times 10^{-4}$, relatively bright for a debris disc, and very similar to
the brightness of the disc around Fomalhaut itself ($f=8 \times 10^{-5}$). For our
stellar luminosity, assuming grains that absorb and emit as blackbodies, this temperature
corresponds to dust at a radial distance of about 10 AU.

Though uncertain, the disc diameter implied by blackbody grains is therefore 20 AU. The disc is probably larger than this estimate however; low-mass stars
such as LP~876-10 and GJ~581 are unable to remove small grains with radiation pressure
\citep[e.g.][]{2012A&A...548A..86L}, and due to inefficient long-wavelength emission
small grains are hotter than blackbodies. The ratio of actual to blackbody size,
$\Gamma$, has been derived for many discs, and varies from near unity for early A-type
stars, to values of 3-4 for Sun-like stars
\citep[e.g.][]{2012ApJ...745..147R,2013MNRAS.428.1263B,2013ApJ...776..111M}. The paucity
of debris discs around M-types means that this ratio has only been measured for GJ~581
\citep[$\Gamma=9$,][]{2012A&A...548A..86L} and AU Mic \citep[$\Gamma \approx
3$,][]{2008ApJ...681.1484R,2012ApJ...749L..27W}. For LP~876-10, using a simple face-on
uniformly bright ring model, we found that disc radii larger than about 40 AU were
inconsistent with the images, suggesting a limit of $\Gamma<4$. This simple ratio is of
course subject to uncertainty if the disc does not lie in a narrow ring or the observed
structure varies with wavelength, but with a sufficient number of objects over a wide
range of stellar luminosities it will be an important future diagnostic of grain
properties in debris discs.

\begin{figure}
  \begin{center}
    \hspace{-0.25cm} \includegraphics[width=0.48\textwidth]{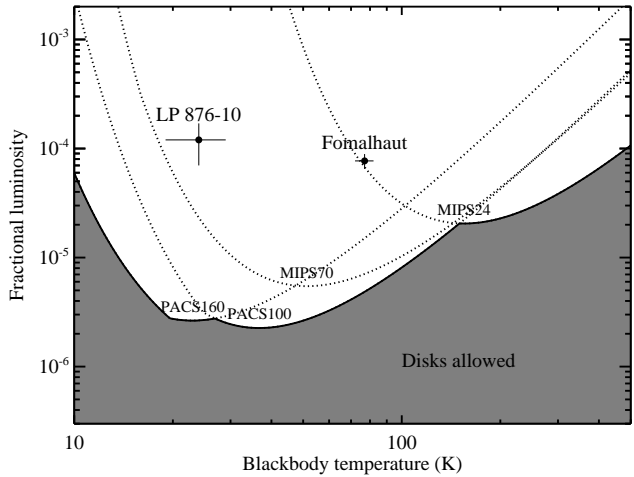}
    \caption{Disc detection limits for TW~PsA (Fomalhaut~B). Dotted lines show upper
      detection limits for discs from different observations (as labelled), and the grey
      region shows where discs are allowed given the non-detection of excess by all
      observations. Dots show the locations of the Fomalhaut and LP~876-10 (Fomalhaut~C)
      blackbody disc models and 1$\sigma$ uncertainties. Because they lie in the white
      region, discs with these temperatures and fractional luminosities would have been
      easily detected around TW~PsA.}\label{fig:lim}
  \end{center}
\end{figure}

\subsection{TW~PsA (Fomalhaut B)}

TW PsA was observed in December 2010. The images have relatively low S/N, but appear
consistent with unresolved emission. We obtained fluxes of $10.1 \pm 1.9$ mJy and $7.2 \pm
3.8$ mJy at 100 and 160 $\mu$m respectively. The best fitting photospheric model has
$T_{\rm eff}=4600 \pm 10 $K and $L_\star=0.187 \pm 0.002 L_\odot$, with photospheric
fluxes of 12.8 and 4.9 mJy at 100 and 160 $\mu$m respectively. The observed fluxes are not
significantly different from the photospheric values, so no excess is detected and we did
not observe this target again. Figure \ref{fig:lim} shows the limits on the fractional
luminosity of a disc around TW~PsA given these observations, combined with limits also
set at 24 and 70 $\mu$m by the Multiband Imaging Photometer for \emph{Spitzer}
\citep[MIPS,][]{2008ApJS..179..423C}. Discs as bright as those around Fomalhaut and
LP~876-10 would easily have been detected around TW~PsA, and any disc around TW~PsA that
has a similar temperature must be more than an order of magnitude fainter than those
discs.

\subsection{Confusion?}

It is possible that the observed \emph{Herschel} excess emission arises from chance
alignment of LP~876-10 with a background galaxy. The chance of a 15.5 mJy or brighter
background source appearing within 5\farcs0 (the \emph{Herschel} pointing accuracy is
about 2\farcs0 at 1$\sigma$) at 160 $\mu$m is about 1\% \citep{2013MNRAS.428L...6S}. This
estimate is conservative; a background galaxy randomly placed within a 5 arcsec radius of
some position is more likely to be at the outer edge of this circle, and the emission was
twice observed to be within 2 arcsec of the expected position.

Given the 89 M dwarfs observed by DEBRIS, one would expect $\sim$1 case of background
galaxy alignment among these stars. However, that LP~876-10 is among the M dwarfs
observed by DEBRIS is a red-herring for the confusion estimate; this discovery could have
been made by anyone with an interest in LP~876-10 and access to the \emph{Herschel}
archive, in which case the DEBRIS sample would not have been considered for the confusion
analysis. Of course, LP~876-10 is also among the sample of all stars that were observed
by \emph{Herschel} (i.e. background confusion is independent of stellar properties and
observing programmes). This study was motivated by the recent discovery by
\citet{2013AJ....146..154M} that LP~876-10 has a kinematic association with Fomalhaut, a
property that, aside from the age of the Fomalhaut system, is independent of the presence
of a debris disc. Therefore, we estimate a $\sim$1\% chance of confusion of LP~876-10.

Another consideration is of system plausibility, in essence consideration of priors that
influence the likelihood of the LP~876-10 excess being due to a debris disc. The main
property of interest is stellar age; if LP~876-10 were a field M dwarf with a random age
between 0-10 Gyr, the chance of a disc detection would be low, but given the prevalence
of debris around young stars, the association of this star with 440 Myr-old Fomalhaut
makes the chance of disc detection more likely and less surprising. Considering binarity,
the eccentricity of the debris ring and planet around Fomalhaut itself are both
suggestive of possible interactions with another star, for which LP~876-10 is a probable
candidate. Such an interaction would also stir up material around LP~876-10, again making
the detection of a disc more likely.

Therefore, while follow-up observations are warranted, we conclude based on the above
discussion that the LP~876-10 excess is most likely associated with the star itself, and
interpret this emission as arising from dust in a debris disc.

\section{Discussion}

The discovery of a debris disc around LP~876-10 is remarkable for several reasons; i)
debris discs are rarely detected around late-type stars, ii) the Fomalhaut system is only
the second case where two debris discs have been found in a stellar system, and iii) the
primary is host to both a bright debris disc and an exoplanet whose eccentricities may be
related to interaction(s) with one or both of the wide companions. We now briefly explore
each of these aspects.

Currently, the lowest mass nearby star thought to host a debris disc is the M4V star
AT~Mic, a member of the $\beta$ Pictoris Moving Group
\citep{2001ApJ...562L..87Z,2009ApJ...698.1068P}. Excesses have also been detected for
low-mass stars in young clusters \citep[e.g. M5V ID10 in the 154 pc distant, 50 Myr old
IC 2391,][]{2007ApJ...654..580S}, though these systems are less well characterised due to
their greater distance. We attempted to verify the AT~Mic 24 $\mu$m excess reported by
\citet{2009ApJ...698.1068P}, but found a greater photospheric 24 $\mu$m flux (144
vs. 114 mJy). This higher flux agrees with the observed value (when colour corrected for
the stellar spectrum) of 143 mJy, and thus we find no IR excess for AT~Mic. The most
likely reason for the difference is that we used the more recent AMES-Cond PHOENIX models
\citep{2005ESASP.576..565B}, though we also included more mid-IR photometry (from WISE
and AKARI) in fitting our atmosphere model. Given difficulties with modelling M dwarf
photospheres, and a lack of excess detection by \citet{2008ApJ...681.1484R} and
Riviere-Marichalar et al. (A\&A submitted), we do not consider the excess for this star
to be robust. Thus, LP~876-10, also with a spectral type of M4V, is one of the lowest
mass stars thought to host a debris disc, and as a bright nearby star will be an
important object for future M dwarf debris disc work.

While debris discs are rarely detected around late-type stars, the most probable reason
is that due to the low luminosity of their host stars, their discs are cool and faint and
only emit significantly at far-IR wavelengths. Sensitivity at these wavelengths is poor
relative to the photospheric level, so only the brightest discs can be detected
\citep[e.g. see][figure 2]{2009ApJ...698.1068P}. In addition, the LP~876-10 disc does not
emit strongly at 70 $\mu$m, so would probably not have been detected even if it had been
observed at this wavelength by \emph{Spitzer}. With $\sim$100 late-type stars observed by
\emph{Spitzer} and \emph{Herschel}, the disc around LP~876-10 therefore lies among the
brightest few percent of M dwarf debris discs. The fractional luminosity of $f=1.2 \times
10^{-4}$ is not extreme enough to exclude the standard picture of a debris disc born at
the same time as the star however \citep[e.g.][]{2008ARA&A..46..339W}. At the system age
of 440 Myr \citep{2012ApJ...754L..20M} collisional grinding is expected to have reduced
the disc mass and luminosity, with fractional luminosities above roughly $10^{-4}$ to
$10^{-3}$ excluded by collisional mass loss for this disc radius and age
\citep{2007ApJ...658..569W}. Thus, while the disc may have originated in a protoplanetary
disc that was more massive than average, a scenario where the disc was created more
recently is not required (but as noted below is of course possible).


Of the three stars in the Fomalhaut system, both the primary and tertiary are now known
to have debris discs. Debris discs around multiple components of multiple stellar systems
are rare, with the only other example being the HD 223352 system, with a circumbinary disc around an A0V +
K-type close binary, and a second disc around the third object, the K1V star HD 223340
\citep{PhillipsThesis}. At 75 arcsec from the primary and with a system distance of 42
pc, HD 223340 is at least 3000 AU from the primary. The Fomalhaut system is even more
loosely bound, with LP~876-10 at 0.77 pc from the primary \citep[158
kAU,][]{2013AJ....146..154M}. This rarity is perhaps surprising given that multiple
stellar systems with multiple protoplanetary discs are relatively common
\citep[e.g.][]{2007prpl.conf..395M}, though may in part be due to the later companion
spectral types.

The detection of two discs in these systems might be attributed to their young ages;
debris discs are observed to become fainter over time
\citep[e.g.][]{2003ApJ...598..636D}, a result of collisional evolution
\citep[e.g.][]{2003ApJ...598..626D,2007ApJ...663..365W}, so their discovery is more
likely if the host stars are young. HD 223352/40 belongs to the $\sim$70 Myr old AB
Doradus moving group \citep{2011ApJ...732...61Z} so the detection of two discs is perhaps
not surprising. The Fomalhaut system is about 440 Myr old, which is young relative to
randomly selected M-type stars, whose ages will be evenly distributed up to $\sim$10
Gyr. Therefore LP~876-10 is among the youngest $\sim$5\% of M dwarfs, which may explain
the presence of a bright debris disc despite detections generally being rare. Many young
stars are not seen to have debris however, so the lack of a bright disc around TW~PsA may
simply mean that the disc around this star had a smaller radius and has evolved to
obscurity by more rapid collisional evolution, or that it was initially lower in mass
than the other two.

Considering how the LP~876-10 debris disc fits within the greater context of the
Fomalhaut system, the properties of note are the large separations of TW~PsA and
LP~876-10, the spectacular offset debris ring \citep{2005Natur.435.1067K}, and the high
eccentricity planet, Fomalhaut b \citep{2013ApJ...775...56K}. That LP~876-10 lies along
the same position angle as the Fomalhaut disc major axis, which is only $24^\circ$ from
edge-on, may be a hint that both currently orbit in the same plane
\citep{2013ApJ...775...56K}, though the LP~876-10 plane will vary over time due to
Galactic tides (see below). A key question is whether the companions ever come very close
to the primary (i.e. have small pericenter distances) or each other, which could
influence the observed properties of the debris discs. There seem to be two formation
scenarios for this wide multiple system, each with differences in the expected companion
orbital evolution; relatively gentle multiple system formation during cluster evaporation
\citep[e.g.][]{2010MNRAS.404.1835K,2011MNRAS.415.1179M}, and dynamical interactions in a
triple system where hardening of an inner binary results in a wider companion
\citep[e.g.][]{2012Natur.492..221R}.

If the Fomalhaut system was originally more compact overall and LP~876-10 dynamically
thrown to a much larger orbit, it would have a highly eccentric orbit and the two debris
discs could be related. However, this formation scenario relies on the hardening of an
inner binary, and TW~PsA is itself probably too well separated from Fomalhaut to be the
cause \citep{2012Natur.492..221R}. It is also very unlikely that Fomalhaut is itself a
close binary \citep[e.g.][]{2013ApJ...764....7K} so the inner binary hardening scenario
seems unlikely here.

If the Fomalhaut system formed essentially by chance when momentarily bound stars in a
cluster became a permanent system, then their inclinations should be random, and their
eccentricities are expected to be weighted towards higher values \citep[a median $e$ of
about 0.7, ][]{2010MNRAS.404.1835K}. It is therefore unlikely, though possible, that
either TW~PsA or LP~876-10 would initially have an eccentricity sufficient to come very
close to Fomalhaut (within 500 AU, or $e>0.98$, say). This eccentricity may evolve
however, due to interactions between the companions \citep[e.g.][]{1962AJ.....67..591K},
and/or the effect of Galactic tides \citep[e.g.][]{1986Icar...65...13H}. The timescale
for the eccentricity of TW~PsA to vary due to Kozai cycles depends on the companions'
orbits, which are unknown, but for circular orbits at their known separation is $\sim$2.7
Gyr \citep[e.g.][]{2006A&A...446..137B}, meaning that this mechanism may not have had
time to cause this companion to interact strongly with the primary unless the initial
eccentricity was already very high. Such a scenario also provides no link between the
Fomalhaut and LP~876-10 discs. Galactic tides however, become stronger for wider orbits,
and the timescale estimate given by \citet{1986Icar...65...13H} suggests that the
companion orbits will vary on timescales similar to the orbital periods ($\sim$10-100
Myr). With the caveat that the system must also remain stable for the system age of
$\sim$440 Myr, the effects of Galactic tides may dominate the dynamics of the companions,
and thus may also have lead to sufficiently close encounters of either companion with the
primary \citep[see also][]{2013Natur.493..381K}.

Whether there is any link between the discs around Fomalhaut and LP~876-10 is unclear. If
the stars have always remained well separated, the evolution of the two bright debris
discs should be no different to random single stars, and their detection in the Fomalhaut
system would be attributed to the relative youth of the system (in particular
LP~876-10). Alternatively, the wide separation of the companions may lead to complex
dynamics; the bright debris discs and eccentric planet may have a common cause due to a
past interaction between Fomalhaut and LP~876-10, which stirred up their debris discs,
perhaps igniting a collisional cascade in a previously quiescent disc
\citep[e.g.][]{2002AJ....123.1757K}, or provoking an instability in the planetary system
\citep[e.g.][]{2011MNRAS.411..859M} that later stirs the disc. Such scenarios are of
course speculation, but motivate detailed observations of all system components. It may
be that high resolution observations of the LP~876-10 disc will reveal evidence of
dynamical perturbations, a signature of yet another planet (e.g. Fomalhaut Cb) or a
highly eccentric stellar orbit that periodically brings it closer to Fomalhaut or TW~PsA.

\section*{Acknowledgements}

We thank the referee for a valuable report. This work was supported by the European Union
through ERC grant number 279973 (GMK \& MCW). P.K. acknowledges support from NASA
NNX11AD21G, NSF AST-0909188 and JPL/NASA award NMO711043.


\end{document}